\documentclass[runningheads]{llncs}
\pdfoutput=1
\usepackage{graphicx}
\usepackage{hyperref}

\usepackage[misc]{ifsym}
\usepackage{amsmath}
\usepackage{amssymb}
\usepackage{subcaption}
\usepackage{ifthen}

\usepackage{xspace}

\newcommand*{\ie}{i.\,e.\xspace}
\newcommand*{\Iltis}{\textsc{Iltis}\xspace}

\usepackage{tikz}
\usetikzlibrary{shapes.misc}
\usetikzlibrary{decorations.text,decorations.footprints,decorations.shapes,shapes.symbols}
\pgfdeclarelayer{background}
\pgfdeclarelayer{foreground}
\pgfsetlayers{background,main,foreground}
\usetikzlibrary{matrix,calc,positioning,backgrounds,fit}
\usetikzlibrary{shapes.multipart,shadows,shadows.blur}

\newcommand{\Mstrut}{\vrule height 0.4cm depth 0.1cm width 0pt}
\newcommand{\taskinput}[1]{\textbf{Input:} #1}
\newcommand{\taskoutput}[1]{\textbf{Output:} #1}
\newcommand{\taskcontent}[3]{\vspace{-1ex}\ifthenelse{\equal{#1}{}}{}{\textbf{#1: }}\textbf{#2}\ifthenelse{\equal{#3}{}}{}{\\[-1ex] \Mstrut #3}\vspace{-1ex}}
\newcommand{\smalltaskcontent}[5]{%
    \vspace{-1ex}\ifthenelse{\equal{#1}{}}{}{\textbf{#1: }}\textbf{#2}%
    \ifthenelse{\equal{#3}{}}{}{\\[-1.3ex]\textbf{\Mstrut #3}}%
    \ifthenelse{\equal{#4}{}}{}{\textbf{:}\\[-1ex] \Mstrut #4}%
    \ifthenelse{\equal{#5}{}}{}{\\[-1.3ex]\Mstrut #5}%
    \vspace{-1ex}}

\newcommand{\task}[5]{%
     \nodepart{one} \taskinput{#4}
     \nodepart{two} \taskcontent{#1}{#2}{#3}
     \nodepart{three} \taskoutput{#5}
}
\newcommand{\smalltask}[7]{%
     \nodepart{one} \taskinput{#6}
     \nodepart{two} \smalltaskcontent{#1}{#2}{#3}{#4}{#5}
     \nodepart{three} \taskoutput{#7}
}

\definecolor{iltisheader}{rgb}{1,.89,.635}
\definecolor{iltistask}{rgb}{1,.95,.816}
\definecolor{iltisedge}{rgb}{.945,.725,.333}
\definecolor{iltispopupheader}{rgb}{.75,.675,.607}

\definecolor{iltispurpel3}{rgb}{.522,.471,.651}
\definecolor{iltispurpel4}{rgb}{.435,.373,.588}
\definecolor{iltisblue0}{rgb}{.89,.95,1}
\definecolor{iltisblue1}{rgb}{.8,.91,1}
\definecolor{iltisblue2}{rgb}{.55,.72,.85}
\definecolor{iltisgrey1}{rgb}{.93,.92,.91}
\definecolor{iltisgrey2}{rgb}{.85,.84,.82}
\definecolor{iltisgrey3}{rgb}{.76,.75,.74}

\newcommand{\borderradius}{1pt}
\newcommand{\borderwidth}{.8pt}

\tikzset{
	task/.style={
		rectangle split,
		rectangle split parts=3,
		rectangle split part fill={iltisblue1, iltisblue0, iltisblue1},
		rectangle split part align={left, center,left},
		rectangle split draw splits=false,
        every one node part/.style={font=\scriptsize},
        every two node part/.style={font=\footnotesize},
        every three node part/.style={font=\scriptsize},
		rounded corners=3pt,
		inner xsep=5pt,
		inner ysep=4pt,
		align=center,
		line width=.3pt,
    },
	taskedge/.style={
        ->,
        draw,
        line width = 2pt,
        iltisblue2,
        shorten <=2pt,
        shorten >=2pt
    },
	exercise/.style={
        node distance = 0.3cm,
		every node/.style={task},
		every edge/.append style={taskedge},
	},
    connector/.style={
        decoration={footprints,foot of=felis silvestris,foot length=5pt,stride length=10pt,foot sep=1pt},
        decorate,
        iltisedge,
    },
    comment/.style={
		draw=iltisheader,
		fill=iltistask,
		rectangle,
        rounded corners=\borderradius,
		line width=\borderwidth,
		inner xsep=5pt,
		inner ysep=4pt,
		node font=\scriptsize,
	}
}

\newcommand{\includeScreenshot}[6]{
	\includegraphics
	[%
	width=#2,%
	clip,%
	trim=#3 #4 #5 #6,%
	]%
	{#1}%
}

\tikzset{screenshot/.style={
    inner sep=0,outer sep=0,
    align=center, 
}}

\tikzset{screenshot border/.style={
    draw=iltisheader,
    rounded corners=\borderradius,
    line width=\borderwidth,
    inner sep=0,outer sep=0,
    inner xsep=-1.5pt,
    align=center, 
}}

\usepackage{pgfplots}
\pgfplotsset{compat=1.16}

\usepgfplotslibrary{dateplot}
\usetikzlibrary{pgfplots.dateplot}

\usepackage{csquotes}
\usepackage{tabularx}
\usepackage{array}
\newcolumntype{x}[1]{>{\centering\arraybackslash\hspace{0pt}}p{#1}}
\usepackage{enumitem}
\usepackage{booktabs}

\begin{document}

\title{\Iltis: Learning Logic in the Web}
\author{
    Gaetano Geck\inst{1}\and
    Christine Quenkert\inst{2}\and
    Marko Schmellenkamp\inst{2}$^\text{\Letter}$\and
    Jonas Schmidt\inst{1}\and
    Felix Tschirbs\inst{2}\and
    Fabian Vehlken\inst{2}\and
    Thomas Zeume\inst{2}$^\text{\Letter}$
}
\authorrunning{Geck et al.}
\institute{TU Dortmund\\
    \and Ruhr University Bochum\\
    \email{\{marko.schmellenkamp, thomas.zeume\}@rub.de}
}

\maketitle

\begin{abstract}
	The \Iltis project provides an interactive, web-based system for teaching the foundations of formal methods.
	It is designed with the objective to allow for simple inclusion of new educational tasks; to pipeline such tasks into more complex exercises; and to allow simple inclusion and cascading of feedback mechanisms. Currently, exercises for many typical automated reasoning workflows for propositional logic, modal logic, and some parts of first-order logic are covered.

	Recently, \Iltis has reached a level of maturity where large parts of introductory logic courses can be supplemented with interactive exercises. Sample interactive course material has been designed and used in courses over the last years, many of them with more than 300 students.

	We invite all readers to try out \Iltis: \url{https://iltis.cs.tu-dortmund.de}

\end{abstract}

\section{Introduction and motivation}
Formal, logical foundations are a cornerstone for many applications of computer science, including formal verification, artificial intelligence and database systems. For this reason, recommendations for Bachelor computer science curricula include logical foundations as an integral part (see, e.g., \cite{ACM2013,GI2016}).

Teaching formal foundations of computer science is a challenge for instructors as it is one of the harder topics for many students. Handling diverse backgrounds for a rapidly increasing number of students enrolled in computer science courses \cite{CRA2017} is problematic and providing individual human tutoring for so many students not always possible, in particular in online teaching scenarios, e.g., under COVID-19 conditions or in freely accessible online courses (MOOCs). A general approach for increasing learning outcomes across STEM disciplines is provided by the National Research Council of the US which advocates, among others, to \enquote{Leverage technologies to make the most effective use of students' time, shifting from information delivery to sense-making and practice in class} \cite{Singer2012,Beach2012}.

The \Iltis project aims at implementing this objective for the foundations of formal methods \cite{GeckLPSVZ18,GeckLHMSSSTVZ19}. The project offers a web-based teaching support system for the logic domain, with a flexible, modular framework for designing and combining educational tasks. A variety of educational tasks for typical workflows in formal methods have already been implemented (see Section~\ref{sec:tasks}). For instance, standard reasoning workflows ---  modeling scenarios with logical formulas, transforming formulas into suitable normal forms, and inferring new knowledge --- are covered for propositional logic and modal logic, and partially for first-order logic. For each task, automated feedback is given which can be declaratively configured by combining predefined feedback generators (see Section~\ref{sec:feedback}). The ability to combine small educational tasks into more complex exercises and workflows is a core aspect of \Iltis.

\subsubsection*{Contribution}
The objective of this article is to announce that the \Iltis teaching support system has reached maturity for broad use in introductory courses for formal methods.

As a proof-of-concept, extensive interactive material has been designed for an introductory course \emph{Logic for computer scientists} aimed at introducing basic concepts, algorithms, and theoretical foundations for propositional logic, modal logic and first-order logic to 2nd year computer science students. The interactive material includes exercises for practicing concepts, multiple choice quizzes, and interactive tutorials for each chapter of the course.

The focus of development since the initial announcement of the \Iltis project~\cite{GeckLPSVZ18} has been on the following aspects:

\begin{itemize}
    \item \textbf{Broadening the portfolio of educational tasks:} A focus has been on supporting more standard workflows from formal methods. The available portfolio of tasks can be used flexibly to cover a multitude of typical workflows for propositional logic, modal logic, and first-order logic. This includes, for instance, reasoning workflows for propositional and modal logic (modeling scenarios by logical means; transformation into normal forms; drawing inferences with the tableau calculus, resolution, or other calculi; and with intermediate testing of understanding via multiple choice questions) as well as  workflows for deciding equivalence of formulas (formulating decision hypotheses;
    \item \textbf{Improving the quality and flexibility of feedback:} The feedback system has been re-implemented from scratch to allow instructors to specify how feedback is provided in a flexible, yet simple declarative language.
    \item \textbf{Increasing usability and user experience:} The graphical user interface has been re-designed from scratch, taking didactical and usability aspects into account.
\end{itemize}

\subsubsection*{Related work}
A large variety of teaching support systems for the foundations of formal methods have been developed over the years.  As most of these systems were developed ad-hoc by instructors for helping their students, a common theme is that they only cover a small set of topics (typically: only one) and are
abandoned and/or become technologically outdated rather quickly. For instance, of the 26 tools described by
Huertas \cite{Huertas2011}, at least 13 are not available anymore (as of November 2021) and many of the remaining
tools use outdated technology. There are only few modern, web-based tools:

\begin{itemize}
    \item \emph{For teaching the foundations of logic:} Modeling for propositional logic is supported by the DiMo tool \cite{HundeshagenLS21}. The \emph{LogEX} system supports transformation of propositional formulas \cite{LodderHJ15,LodderH11}. The focus of the \emph{Logic4Fun} project is on modeling scenarios and puzzles by logical means \cite{Slaney15}.
    \emph{AELL} \cite{HuertasHLM11} offers tasks for natural deduction, truth tables, and resolution but is not publicly available. Modal logic semantics can be explored with \emph{Hintikka's World} \cite{CharrierGNS19}.
    \item \emph{For teaching the foundations of formal languages:} Constructing models for formal languages such as regular expressions, automata, and formal grammars can be practiced with a variety of tools. In \emph{FLACI} \cite{HielscherW19}, students can convert between these models. The \emph{AutomataTutor} \cite{AntoniKAGV2015,AntoniHKRW2020} offers elaborate feedback on finite automata construction tasks. \emph{JFLAP} \cite{GramondR1999,Rodger1999} features many algorithms on automata, but is not web-based.
\end{itemize}

\section{Designing educational content in \Iltis}\label{sec:tasks}

Instructors can design educational content flexibly by using the broad portfolio of educational tasks available and the compositionality of such tasks.

\subsubsection*{Compositional task model}
\begin{figure}
    \newlength{\taskwidth}
    \setlength{\taskwidth}{.45\textwidth}
    \newlength{\sswidth}
    \setlength{\sswidth}{.455\textwidth}
    \newlength{\commentwidth}
    \setlength{\commentwidth}{\sswidth}
    \addtolength{\commentwidth}{-2.7ex}
    \begin{tikzpicture}[
            screenshot/.append style={node distance=1mm and 1mm},
            comment/.append style={node distance=1mm and 11mm},
            task/.append style={minimum width=\taskwidth,node distance=7.1mm and 1mm},
        ]

        \node (assignment) [task] {\task{}{Assignment}{}{--}{--}};

        \node (model1) [task,below=of assignment] {\task{Step 1}{Constructing formulas}{Model the scenario}{--}{Formulas $\psi_1, \ldots, \psi_m$}};

        \node (model2) [task,below=of model1] {\task{Step 2}{Constructing formulas}{Model the consequence}{--}{Formula $\psi$}};

        \node (what) [task,below=of model2] {\task{Step 3}{Multiple choice}{What to do now?}{--}{--}};

        \node (transformation) [task,below=of what] {\task{Step 4}{Transforming formulas}{Transform into implication form}{$\psi_1 \wedge \ldots \wedge \psi_m\wedge \neg \psi$}{Formula $\varphi$ in implication form}};

        \node (horn) [task,below=of transformation] {\task{Step 5}{HornSat algorithm}{Apply algorithm}{Formula $\varphi$ in implication form}{--}};

        \node (decide) [task,below=of horn] {\task{Step 6}{Multiple choice}{Determine unsatisfiability}{--}{--}};

        \draw (assignment) edge [taskedge] (model1);
        \draw (model1) edge [taskedge] (model2);
        \draw (model2) edge [taskedge] (what);
        \draw (what) edge [taskedge] (transformation);
        \draw (transformation) edge [taskedge] (horn);
        \draw (horn) edge [taskedge] (decide);

        \node (ss-assignment) [comment,right=of assignment.north east,anchor=north west] {\begin{minipage}{\commentwidth}
            \textbf{Debugging a chat system:}\\
            Archie sends some test messages to his three co-developers Sophie, Luke, and Maja. He makes the following observations: [\ldots\!]
            Archie assumes that Maja was able to receive his message, but is he right?
            Verify Archie's assumption!
        \end{minipage}};
        \node (ss-model1)[screenshot,below=of ss-assignment] {%
            \includeScreenshot{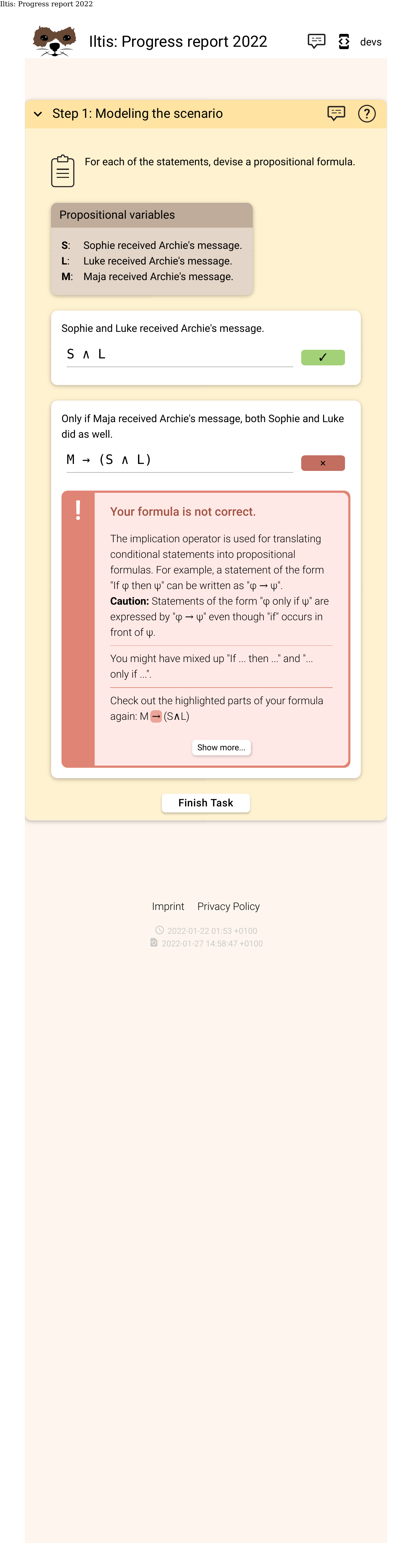}{\sswidth}{13mm}{733mm}{13mm}{52mm}\\[-1pt]%
            \includeScreenshot{images/almodeling_half-feedback_medium.pdf}{\sswidth}{13mm}{382mm}{13mm}{75mm}%
        };
        \node (ss-horn)[screenshot,below=of ss-model1] {%
            \includeScreenshot{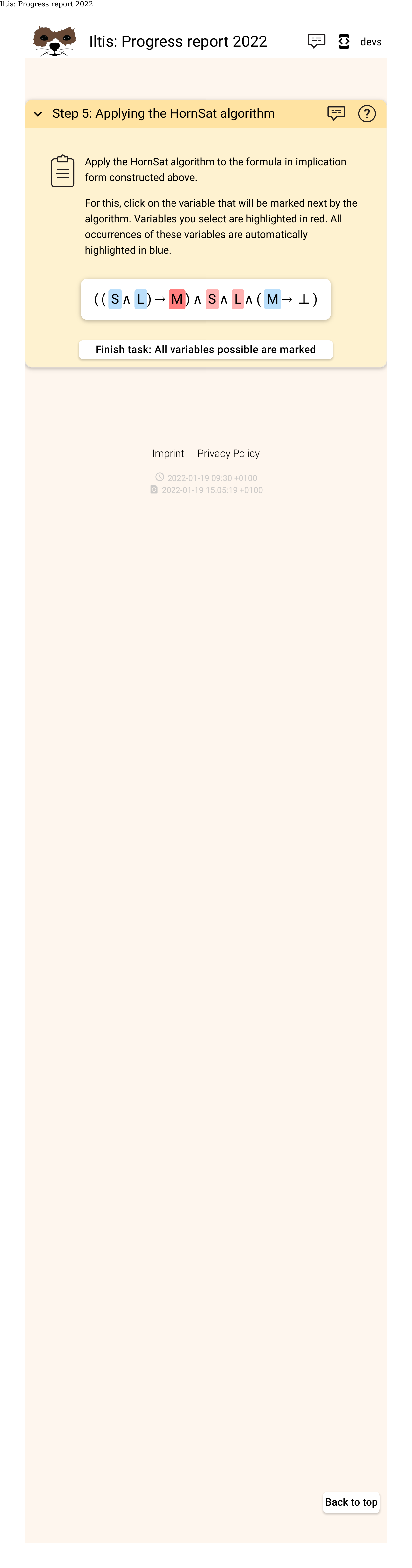}{\sswidth}{13mm}{733mm}{13mm}{52mm}\\[-1pt]%
            \includeScreenshot{images/hornsat_finished_noprotocol_medium.pdf}{\sswidth}{13mm}{614mm}{13mm}{75mm}
        };

        \node (ss-model1)[screenshot border] at (ss-model1) {%
            \phantom{\includeScreenshot{images/almodeling_half-feedback_medium.pdf}{\sswidth}{13mm}{733mm}{13mm}{52mm}}\\[-1pt]%
            \phantom{\includeScreenshot{images/almodeling_half-feedback_medium.pdf}{\sswidth}{13mm}{382mm}{13mm}{75mm}}%
        };
        \node[screenshot border] at (ss-horn) {%
            \phantom{\includeScreenshot{images/hornsat_finished_noprotocol_medium.pdf}{\sswidth}{13mm}{733mm}{13mm}{52mm}}\\[-1pt]%
            \phantom{\includeScreenshot{images/hornsat_finished_noprotocol_medium.pdf}{\sswidth}{13mm}{614mm}{13mm}{75mm}}
        };

        \begin{scope}[on background layer]
            \draw [connector] (assignment.east) -- (ss-assignment.center);
            \draw [connector] (model1.east) -- (ss-model1.center);
            \draw [connector] (horn.east) -- (ss-horn.center);
        \end{scope}
    \end{tikzpicture}

    \caption{An exercise for the complete propositional reasoning workflow, composed of smaller educational tasks. For this sample scenario, the instructor chose the HornSat satisfiability test as Horn formulas are sufficiently expressive. For general propositional formulas, also truth tables, propositional resolution, and the propositional tableau calculus can be used.}\label{fig:alreasoningworkflow}
\end{figure}
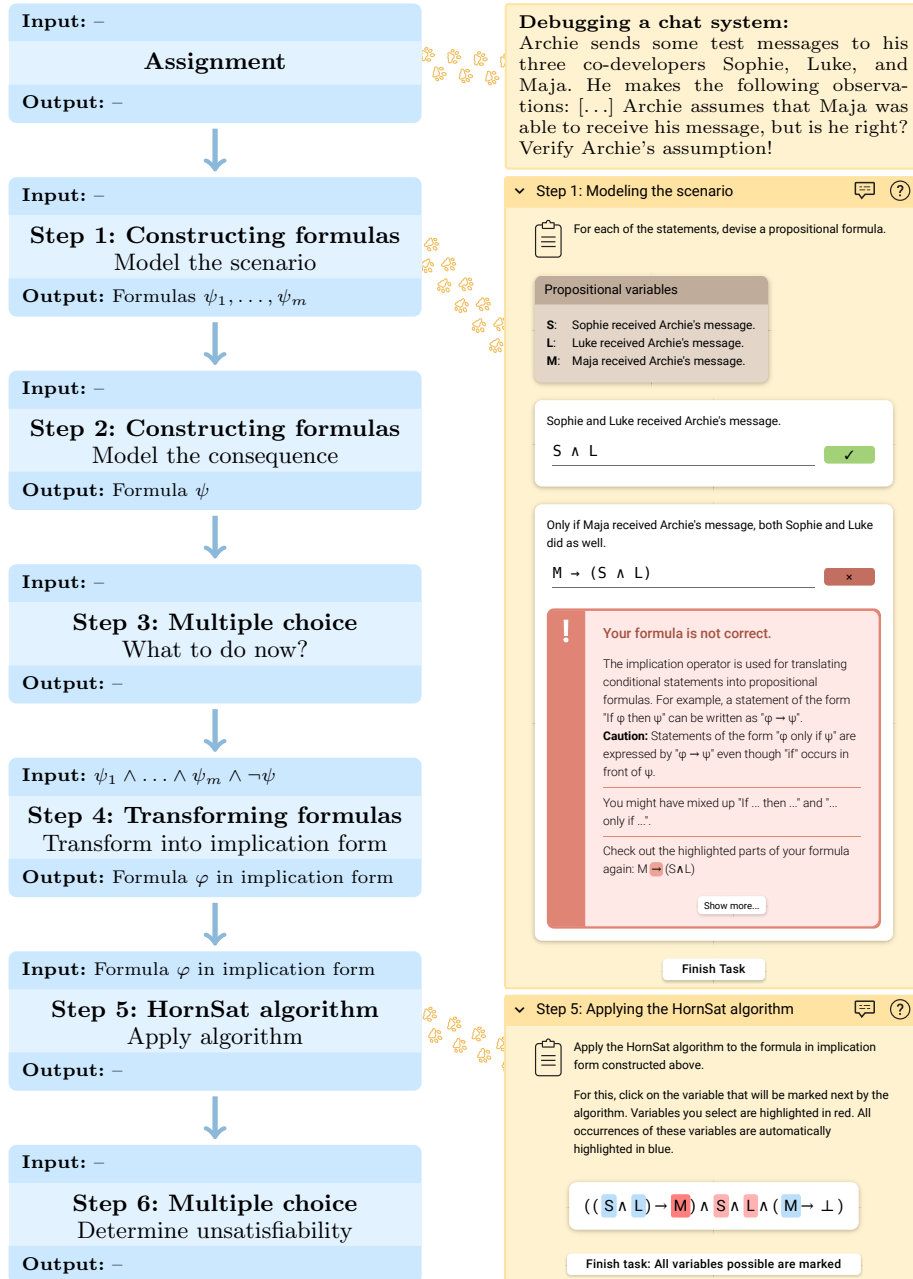

\begin{figure}
    \setlength{\taskwidth}{.44\textwidth}
    \setlength{\sswidth}{.46\textwidth}
    \setlength{\commentwidth}{\sswidth}
    \addtolength{\commentwidth}{-2.7ex}
    \begin{tikzpicture}[
            screenshot/.append style={node distance=1mm and 1mm},
            comment/.append style={node distance=1mm and 1mm},
            task/.append style={minimum width=\taskwidth,node distance=4.5mm and 1mm},
            smalltask/.append style={task,minimum width=.55\taskwidth,node distance=4.5mm and 1mm},
        ]

        \node (assignment) [task] {\task{}{Assignment}{}{--}{--}};

        \node (trafo) [task,below=of assignment] {\task{Step 1}{Transforming formulas}{Transform into NNF}{Formula $\varphi$ (given by teacher)}{Formula $\psi$ in NNF}};

        \node (tableau) [task,below=of trafo] {\task{Step 2}{Tableau Calculus}{Construct a tableau}{Formula $\psi$ in NNF}{--}};

        \node (decide) [task,below=of tableau] {\task{Step 3}{Multiple choice}{Is $\psi$ satisfiable?}{--}{--}};

        \node (kripke) [smalltask,below=of decide,xshift=.3\taskwidth] {\smalltask{4b}{Constructing}{a structure}{Construct a}{model for $\psi$}{Formula $\psi$}{--}};

        \node (whynot) [smalltask,below=of decide,xshift=-.3\taskwidth] {\smalltask{4a}{Multiple}{choice}{Why is $\psi$ not}{satisfiable?}{--}{--}};

        \draw (assignment) edge [taskedge] (trafo);
        \draw (trafo) edge [taskedge] (tableau);
        \draw (tableau) edge [taskedge] (decide);
        \draw (decide) edge [taskedge] (kripke) node [below right=9mm and .15\taskwidth] {$\psi$ satisfiable};
        \draw (decide) edge [taskedge] (whynot) node [below left=9mm and .15\taskwidth] {$\psi$ unsatisfiable};

        \node (ss-assignment) [comment,right=8mm of assignment.north east,anchor=north west] {\begin{minipage}{\commentwidth}
            \textbf{Testing for satisfiability}\\
            Test whether the modal formula
            \vspace{-2ex}
            \[\varphi=\lnot \Box (\lnot A \lor \lnot \Diamond B) \land (\Diamond B \lor \lnot \Diamond A)\]
            \par\vspace{-2ex}
             is satisfiable. If so, also construct a model.
        \end{minipage}};
        \node (ss-trafo)[screenshot,below=of ss-assignment] {%
            \includeScreenshot{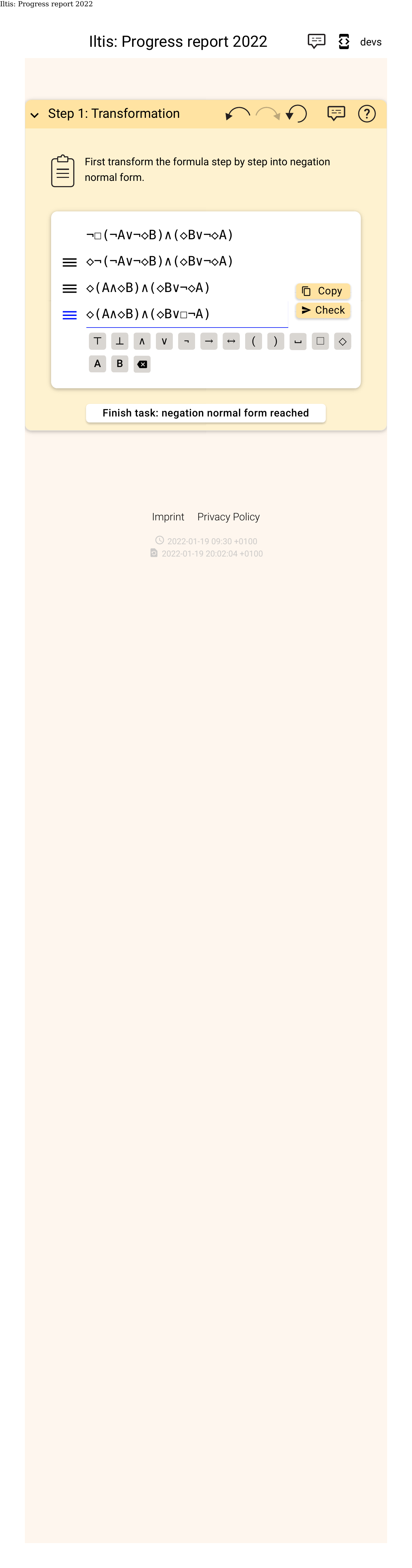}{\sswidth}{13mm}{733mm}{13mm}{52mm}\\[-1pt]%
            \includeScreenshot{images/mlsat-trafo_finished_keyboard_medium.pdf}{\sswidth}{13mm}{581mm}{13mm}{75mm}%
        };
        \node (ss-tableau)[screenshot,below=of ss-trafo] {%
            \includeScreenshot{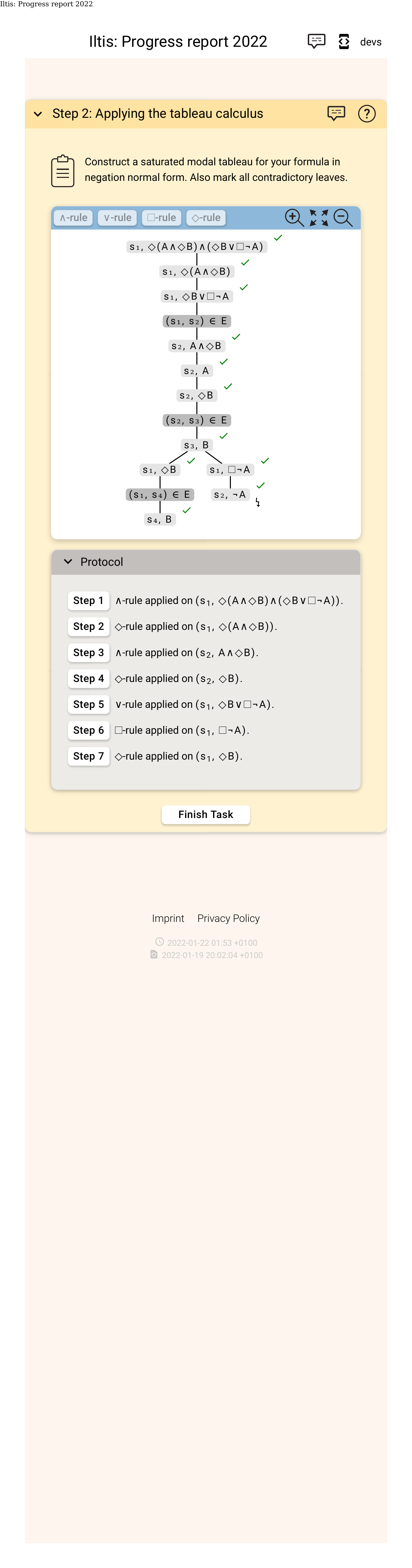}{\sswidth}{13mm}{733mm}{13mm}{52mm}\\[-1pt]%
            \includeScreenshot{images/mlsat-tableau_finished_medium.pdf}{\sswidth}{13mm}{376mm}{13mm}{75mm}%
        };
        \node (ss-kripke)[screenshot,left=6mm of ss-tableau.south west,anchor=south east] {%
            \includeScreenshot{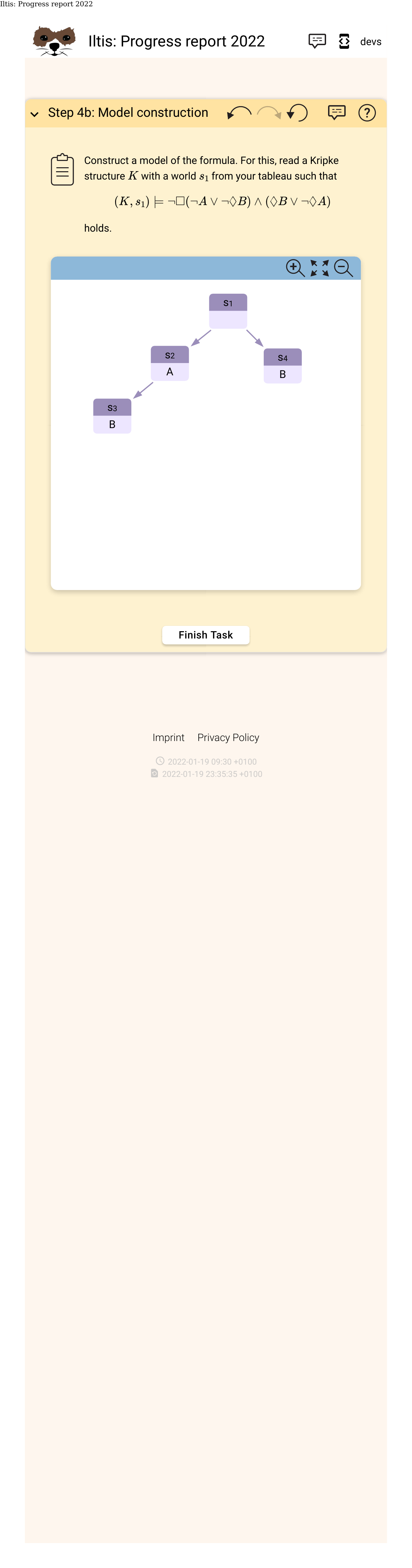}{\sswidth}{13mm}{733mm}{13mm}{52mm}\\[-1pt]%
            \includeScreenshot{images/mlsat-kripke_finished_medium.pdf}{\sswidth}{13mm}{577mm}{13mm}{75mm}\\[-1pt]%
            \includeScreenshot{images/mlsat-kripke_finished_medium.pdf}{\sswidth}{13mm}{493mm}{13mm}{294mm}\\[-2pt]%
            \includeScreenshot{images/mlsat-kripke_finished_medium.pdf}{\sswidth}{13mm}{468mm}{13mm}{318mm}%
        };

        \node [screenshot border] at (ss-trafo) {%
            \phantom{\includeScreenshot{images/mlsat-trafo_finished_keyboard_medium.pdf}{\sswidth}{13mm}{733mm}{13mm}{52mm}}\\[-1pt]%
            \phantom{\includeScreenshot{images/mlsat-trafo_finished_keyboard_medium.pdf}{\sswidth}{13mm}{581mm}{13mm}{75mm}}%
        };
        \node [screenshot border] at (ss-tableau) {%
            \phantom{\includeScreenshot{images/mlsat-tableau_finished_medium.pdf}{\sswidth}{13mm}{733mm}{13mm}{52mm}}\\[-1pt]%
            \phantom{\includeScreenshot{images/mlsat-tableau_finished_medium.pdf}{\sswidth}{13mm}{376mm}{13mm}{75mm}}%
        };
        \node [screenshot border] at (ss-kripke) {%
            \phantom{\includeScreenshot{images/mlsat-kripke_finished_medium.pdf}{\sswidth}{13mm}{733mm}{13mm}{52mm}}\\[-1pt]%
            \phantom{\includeScreenshot{images/mlsat-kripke_finished_medium.pdf}{\sswidth}{13mm}{577mm}{13mm}{75mm}}\\[-1pt]%
            \phantom{\includeScreenshot{images/mlsat-kripke_finished_medium.pdf}{\sswidth}{13mm}{493mm}{13mm}{294mm}}\\[-2pt]%
            \phantom{\includeScreenshot{images/mlsat-kripke_finished_medium.pdf}{\sswidth}{13mm}{468mm}{13mm}{318mm}}%
        };

        \begin{scope}[on background layer]
            \draw [connector] (assignment.east) -- (ss-assignment.center);
            \draw [connector] (trafo.east) -- (ss-trafo.center);
            \draw [connector] (tableau.east) -- ($(ss-tableau.center)-(1cm,0cm)$);
            \draw [connector] (kripke.south) -- (ss-kripke.center);
        \end{scope}
    \end{tikzpicture}
    \caption{An exercise for solving the satisfiability problem for modal formulas, composed of smaller educational tasks. For satisfiable and unsatisfiable formulas, different workflows can be used.}\label{fig:mlsatworkflow}
\end{figure}

Exercises in \Iltis are built from small, easily composable, educational tasks. Each educational task is configured by inputs --- either given explicitly or as the output of prior tasks --- and provides  the objects created by students in this task as outputs, which can then be used by subsequent educational tasks. For instance, a task for transforming a formula into conjunctive normal (CNF) form receives a formula as input and provides the student-constructed, equivalent formula in CNF as output. Two typical multi-step exercises --- a reasoning workflow for propositional logic as well as a satisfiability workflow for modal formulas --- are illustrated in Figures~\ref{fig:alreasoningworkflow} and~\ref{fig:mlsatworkflow} (instantiations of these workflows can be found in the course material\footnote{The course material is accessible at \url{https://iltis.cs.tu-dortmund.de}.}).

\begin{table}[t]
    \newcommand{\gear}[6]{%
      (0:#2)
      \foreach \i [evaluate=\i as \n using {\i-1)*360/#1}] in {1,...,#1}{%
        arc (\n:\n+#4:#2) {[rounded corners=1.5pt] -- (\n+#4+#5:#3)
        arc (\n+#4+#5:\n+360/#1-#5:#3)} --  (\n+360/#1:#2)
      }%
      (0,0) circle[radius=#6]
    }
    \newcommand*{\nosupport}{%
        \raisebox{-1pt}{\scalebox{0.05}{%
        \begin{tikzpicture}%
            \fill[even odd rule] \gear{9}{2}{2.8}{10}{10}{0.8};
        \end{tikzpicture}%
        }}%
    }
    \newcommand*{\supported}{$\checkmark$}
    \newcommand*{\basicsupport}{$(\checkmark)$}
    \newcommand*{\newFeatureLabel}{$^\bullet$}
    \newcommand*{\extendedFeatureLabel}{$^\circ$}
    \newcommand*{\newFeature}{\makebox[1ex]{\newFeatureLabel}}
    \newcommand*{\extendedFeature}{\makebox[1ex]{\extendedFeatureLabel}}
    \newcommand*{\oldFeature}{\phantom{\newFeature}}
    {
    \scriptsize
        \begin{tabularx}{\linewidth}{Xx{2.3cm}x{2.3cm}x{2.3cm}x{2.1cm}}
            \toprule
            Task & Propo\-si\-tio\-nal logic & Modal logic & First-order logic & General \\ \midrule
            Evaluating formulas & \centering \supported\newFeature & \supported\newFeature & \nosupport\oldFeature \\ \midrule[0.05pt]
            Constructing models & \supported\newFeature & \supported\newFeature & \basicsupport\newFeature \\ \midrule[0.05pt]
            Constructing formulas & \supported\oldFeature & \supported\newFeature  & \basicsupport\newFeature \\ \midrule[0.05pt]
            Transforming &  \supported\oldFeature & \supported\newFeature & \nosupport\oldFeature \\ \midrule[0.05pt]
            Testing satisfiability & \supported\extendedFeature & \supported\newFeature & \basicsupport\newFeature
            \\\midrule
            Task variants \newline\& further tasks

            & \raggedright
			    Satisfiability tests with \vspace{-3mm}
				\begin{itemize}[leftmargin=4mm]
				 \item truth tables\newFeatureLabel
				 \item HornSat algorithm\newFeatureLabel
				 \item tableau calculus\newFeatureLabel
				 \item resolution
				\end{itemize}\vspace{-1mm}
                Choosing propositional variables
			& \raggedright
               Satisfiability test with tableau calculus\newFeatureLabel \smallskip

				Calculating bisimulations\newFeatureLabel \smallskip

				Proving non-bisi\-milarity of worlds\newFeatureLabel
			& \raggedright
              Satisfiability test with resolution\newFeatureLabel\smallskip

			  Proving non-equivalence of formulas\newFeatureLabel
			& {\raggedright
              Multiple choice \smallskip

              Messaging\smallskip

              }
			  \\\bottomrule
        \end{tabularx}
	}
    \caption{A summary of educational tasks that are supported (\supported), supported with limitations (\basicsupport), and in development (\protect\nosupport). Functionality that has been added (\newFeatureLabel) or substantially extended (\extendedFeatureLabel) compared to \cite{GeckLPSVZ18} is indicated.}
    \label{tab:tasks}
\end{table}

\subsubsection*{Educational tasks}
\Iltis supports a variety of educational tasks for propositional logic, modal logic, and first-order logic (see Table~\ref{tab:tasks} for a high-level summary):
\begin{itemize}
    \item \emph{Evaluating formulas:} Students can evaluate formulas for a given interpretation. For propositional logic, students can construct truth tables; for modal logic, students can construct evaluation tables for a given Kripke structure.
    \item \emph{Constructing models:} Students can construct models (\ie satisfying interpretations) for formulas. For propositional logic, models are specified by valuations of all variables, for modal logic by Kripke structures. For first-order logic, students can construct models for formulas describing properties of (colored) graphs.
    \item \emph{Constructing formulas:} Students can construct formulas for the respective logics. For propositional and modal logic, formulas are checked for correctness and, if they are not, underlying misconceptions are determined and used to provide adequate feedback (see Section \ref{sec:feedback}). For first-order logic, another approach is taken due to the intractability of equivalence tests: instructors provide a textual description of a unary graph query as well as a sample graph, and students are asked to provide a first-order formula that selects the same nodes as the graph query on this sample graph (see Figure \ref{fig:fomodelling}).
    \item \emph{Applying equivalence transformations:} Students can transform formulas step by step either into an equivalent target formula or into an equivalent formula in a normal form such as conjunctive, disjunctive, or negation normal form. Currently, this educational task is available for propositional and modal logic.
    \item \emph{Testing satisfiability:} Students can test formulas for satisfiability using several methods. For propositional logic, students can a) construct truth tables, b) construct tableaux according to the tableau calculus, c) apply the algorithm for testing satisfiability of Horn formulas (HornSat), or d) construct a resolution graph. For modal logic, students can apply the tableau calculus. For first-order logic, students can construct a resolution graph.
    \item \emph{Determining (non-)bisimilar worlds of Kripke structures:} Students can prove the (non-)bisimilarity by interactively calculating the maximal bisimulation between two Kripke structures using the standard algorithm. Additionally, students can prove non-bisimilarity of two worlds by providing a modal formula that distinguishes them.
\end{itemize}

Further educational tasks for smoothing complex exercise workflows (such as a multiple choice task) and for specific content (such as providing witnesses for non-equivalence) are also available.

\section{Feedback in \Iltis}\label{sec:feedback}
One of the core objectives of \Iltis is to provide immediate and comprehensive feedback, as this is one of the most important factors for learning success.

\subsubsection*{Compositional feedback model}
In the feedback framework of \Iltis, each educational task type comes with multiple \emph{feedback generators}, each one responsible for one kind of feedback. Individual feedback generators can be composed to \emph{feedback strategies} by simple rule-based programs. Such programs determine the order of feedback with the invocation of certain rules possibly depending on the success of previous rules.

When specifying interactive exercises, instructors can state which feedback strategy to use (or define a custom one). In this way, the progress  of students can be taken into account, e.g., a strategy that provides a lot of feedback can be used for beginners, while a strategy that provides almost no feedback can be used for exam preparation. As an example, each section in the feedback box (indicated by thin red lines) in Step 1 of Figure~\ref{fig:alreasoningworkflow} is provided by a different feedback generator.
By step-wise uncovering the feedback of the different generators, students can choose how much of the provided feedback they want to use.

\subsubsection*{Sample feedback strategies and generators}
Designing feedback strategies and generators is a subtle and challenging task as algorithmic feasibility as well as didactical aspects have to be taken into account. We sketch two sample strategies and their feedback generators that are currently provided out-of-the-box for instructors.

A feedback strategy for providing feedback for the construction of propositional formulas can look like this (it is partially illustrated in Step 1 of Figure~\ref{fig:alreasoningworkflow}):
\begin{enumerate}
	\item[(1)] \emph{Correctness:} Is the constructed formula \emph{correct} or \emph{not correct}?
	\item[(2)] \emph{Misconceptions:} Typical misconceptions (e.g., mixing up the premise and the conclusion of an implication, especially in \enquote{only if}-statements) are identified using an abstract rule framework (see \cite{GeckLPSVZ18}). They are the basis for multiple feedback generators:
		\begin{enumerate}
		 \item \emph{Hint at the misconception} (e.g., \enquote{Do you remember how \enquote{if}- and \enquote{only if}-statements can be expressed in propositional logic?})
		 \item \emph{State the misconception explicitly} (e.g., \enquote{You might have mixed up \enquote{if} and \enquote{only if}.})
		 \item \emph{Point out the precise mistake position}
		\end{enumerate}

    \item[(3)] \emph{Distinguishing model:} A valuation that distinguishes the constructed formula from a correct formula.
\end{enumerate}

\begin{figure}
    \begin{minipage}[t]{0.48\linewidth}
        \vspace{0pt}
        \begin{tikzpicture}
            \node[screenshot] {\includeScreenshot{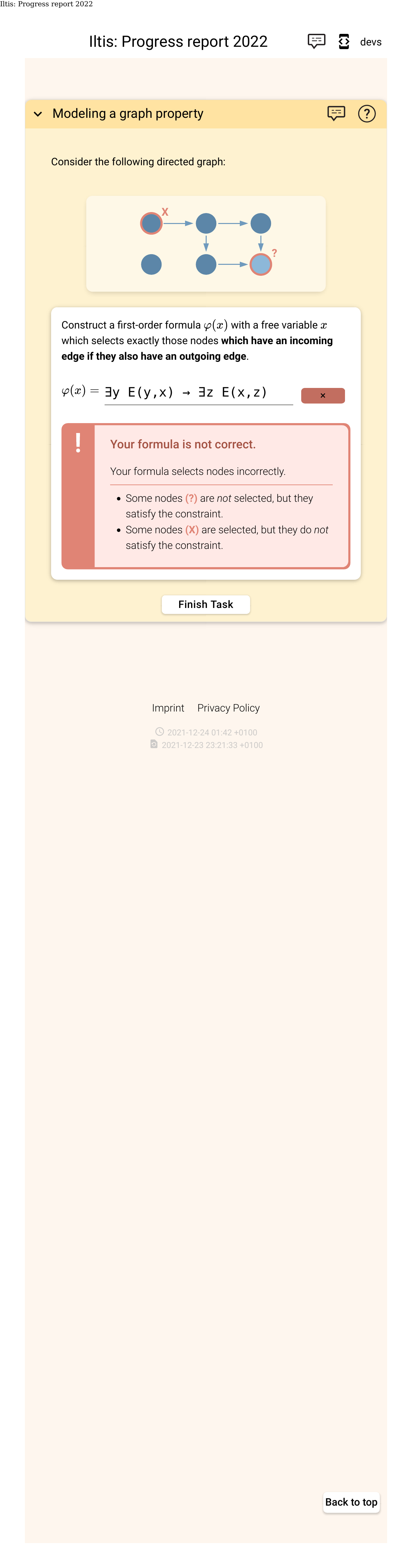}{\linewidth}{13mm}{484mm}{13mm}{52mm}};
            \node[screenshot border]{\phantom{\includeScreenshot{images/fomodelling_feedback_medium.pdf}{\linewidth}{13mm}{484mm}{13mm}{52mm}}};
        \end{tikzpicture}
    \end{minipage}
    \hfill
    \begin{minipage}[t]{0.48\linewidth}
        \vspace{0pt}
        \begin{tikzpicture}
            \node[screenshot] {\includeScreenshot{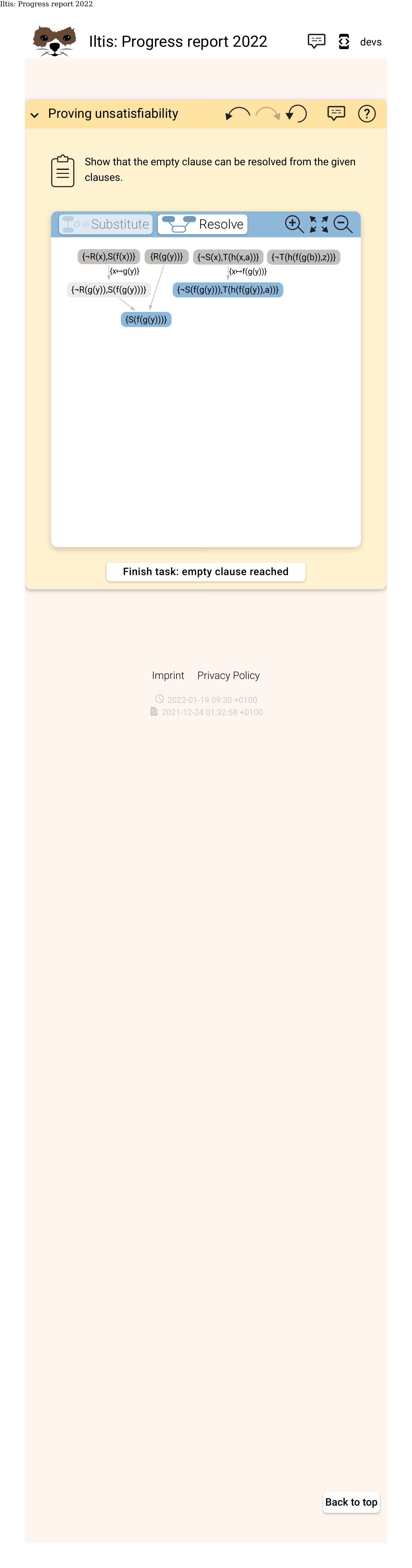}{\linewidth}{13mm}{500mm}{13mm}{52mm}};
            \node[screenshot border] {\phantom{\includeScreenshot{images/foresolution_resolution_nopopup_medium.pdf}{\linewidth}{13mm}{500mm}{13mm}{52mm}}};
            \node[screenshot,
                anchor=north,
                yshift=0,
            ] {\includeScreenshot{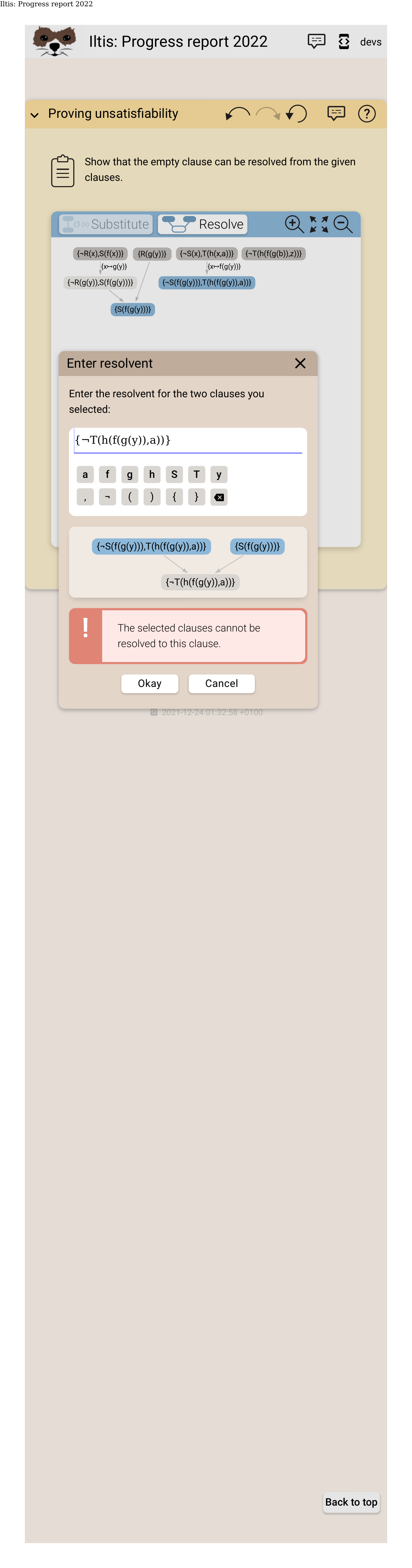}{.6\linewidth}{30mm}{439mm}{48mm}{180mm}};
            \node[screenshot border,
                iltispopupheader,
                anchor=north,
                yshift=0,
            ] {\phantom{\includeScreenshot{images/foresolution_resolution_feedback_medium.pdf}{.6\linewidth}{30mm}{439mm}{48mm}{180mm}}};
        \end{tikzpicture}
        \vspace{0mm}
    \end{minipage}

    \begin{minipage}[t]{0.48\linewidth}
        \caption{An example task for constructing first-order formulas for graph properties. The feedback highlights which nodes of the given graph are erroneously (not) selected by the student-constructed formula.}
        \label{fig:fomodelling}
    \end{minipage}
    \hfill
    \begin{minipage}[t]{0.48\linewidth}
        \caption{An example task for proving unsatisfiability of first-order formulas by constructing a resolution graph. Students can freely apply substitutions and resolution steps until they reach the empty clause.}
        \label{fig:foresolution}
    \end{minipage}
\end{figure}

For providing feedback for the construction of first-order formulas for graph queries, the following feedback strategy can be used. It is shown in action in Figure~\ref{fig:fomodelling}.
\begin{enumerate}
    \item[(1)] \emph{Correctness:} Is the set of nodes selected by the student-constructed formula \emph{correct} or \emph{not correct}?
    \item[(2)] \emph{Subset feedback:} Is the set of nodes selected by the student-constructed formula a subset or superset of the correct set?
    \item[(3)] \emph{Visual feedback:} The difference between the nodes selected by the student-constructed formula and the correct set are highlighted.
\end{enumerate}

Besides construction tasks like the ones explained above, Iltis also supports tasks for training specific step-by-step algorithms. In these learning situations, feedback is not only given at the end of the task, but after each step of the algorithm. This helps students to recognize at which specific step they did not adhere to the algorithm and, thus, to correct their error.
For example, in the task for constructing resolution graphs, every resolution (and possibly substitution) step is checked immediately and feedback is given via a feedback generator dedicated to this step (see Figure~\ref{fig:foresolution}). The other algorithmic tasks (HornSat, Tableau Calculus, Bisimulation, etc.) follow the same principle.

\section{Evaluation of the system}

Interactive course material in \Iltis has been used in an introductory course \enquote{Logic for computer scientists} at TU Dortmund University in the winter terms 2019/2020, 2020/2021, and 2021/2022 as well as at Ruhr University Bochum in the summer terms 2020 and 2021. The use of the material was \emph{voluntary}.

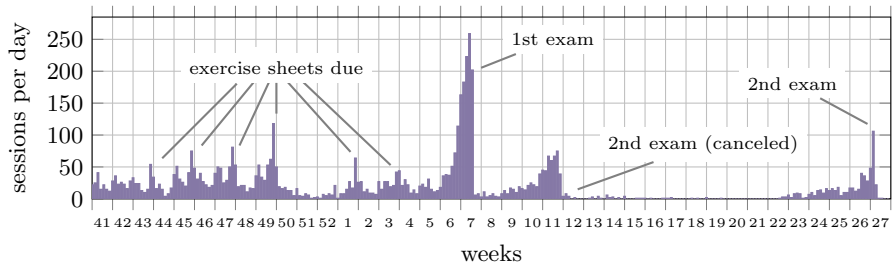
\begin{figure}
    \begin{tikzpicture}[
        every pin/.style={fill=white,font=\scriptsize,pin edge=thick},
        ]
        \begin{axis}[
            ybar,
            date coordinates in=x,
            x tick label as interval,
            bar width=0.7pt,
            x tick label style={font=\tiny},
            y tick label style={font=\footnotesize},
            width=\textwidth,
            height=4cm,
            xlabel=weeks,
            xmajorgrids,
            xmin=2019-10-07,
            xmax=2020-07-06,
            xtick distance=7,
            xticklabels from table={weekLabels_wise2019.csv}{weekLabel},
            ylabel=sessions per day,
            ymajorgrids,
            ytick distance=50,
            ymin=0,
            ]
            \addplot+ [
                iltispurpel3,
                ]
            table [
                x=date,
                y=numberOfSessions,
            ]
            {usagestatistics_wise2019.csv};
            \node[pin={[yshift=-5pt]above right:1st exam}]
                at (axis cs:2020-02-14,200) {};
            \node[pin=above right:2nd exam (canceled)] 
                at (axis cs:2020-03-18,10){};
            \node[pin=above left:{{2nd exam}}] 
                at (axis cs:2020-07-01,110) {};

			\node[pin={[xshift=20pt,yshift=20pt]above right:{}}]
                at (axis cs:2019-10-28,50) {};
			\node[pin={[xshift=13pt,yshift=19pt]above right:{}}]
                at (axis cs:2019-11-11,70) {};
			\node[pin={[xshift=0pt,yshift=17pt]above right:{}}]
                at (axis cs:2019-11-25,70) {};
            \node[pin={[xshift=-21pt,yshift=25pt]above left:{}}]
                at (axis cs:2020-01-06,60) {};
            \node[pin={[xshift=-25pt,yshift=23pt]above left:{}}]
                at (axis cs:2020-01-20,40) {};

            \node[pin=above:{exercise sheets due}]
                at (axis cs:2019-12-09,120) {};
        \end{axis}
    \end{tikzpicture}
    \caption{Access numbers of the \Iltis course website for the logic course at TU Dortmund University in the winter term 2019/2020. These numbers represent the number of sessions per day, with a session starting when a student accesses \Iltis in their browser and ending when this student is inactive for at least 30 minutes. A total of about 380 students actively participated until the end of the course.
    The deadlines for handing in the biweekly exercise sheets and the exams are reflected by peaks in the access numbers.}
    \label{fig:accessnumbers}
\end{figure}

The usage statistics indicate that students rely on \Iltis a lot, in particular for preparing their solutions to biweekly exercise sheets and for their exams (see sample data in Figure \ref{fig:accessnumbers}). Additionally, the courses and also the Iltis system itself were evaluated by the students using non-mandatory questionnaires. Many students stated that \Iltis was very useful for understanding the contents of the lecture and in particular for training methods and algorithms.

We highlight that a controlled empirical study of the effectiveness of \Iltis is far beyond the scope of this paper. Due to fairness required by law, it is hard to set-up such a study with limited resources.

\section{Future work and (shameless) call for help}
In this demonstration, we have described \Iltis, a web-based system offering interactive exercises for learning propositional logic, modal logic, and first-order logic. It is used as e-learning tool accompanying an introductory course on logic covering most of its major topics and can be freely tested at \url{https://iltis.cs.tu-dortmund.de}. In the future, we plan to complete the portfolio of \Iltis and extend it to other foundational areas of computer science.

Our vision for \Iltis is to provide opportunities to re-think approaches for teaching formal foundations. As an example, a technology-assisted introduction to first-order logic as modeling language could look as follows:
\begin{quote}
    After providing intuition by modeling a sample scenario in first-order logic, an instructor asks students to model statements in the same application context in a web system on their mobile devices. The web system analyzes the student attempts and provides the instructor with learning analytics on the distribution of types of mistakes in graphical form, which they can discuss in-class. Afterwards the instructor presents formal syntax and semantics of first-order logic. Before the tutorial sessions, each student individually trains modeling with first-order formulas in different application contexts in the web system, which provides immediate feedback ensuring early correction of misconceptions. At the beginning of the tutorial sessions, most students are able to model with first-order logic and thus larger, more realistic and open-ended modeling tasks can be tackled.
\end{quote}

\noindent To implement this vision, we are looking for:

\begin{itemize}
    \item Instructors who want to use \Iltis in their courses and thereby provide helpful feedback for future directions.
    \item Interested students, PhD students, and PostDocs for helping us to a) lay the theoretical foundations for \Iltis (e.g. the development of suitable feedback mechanisms in algorithmically hard domains) and to b) drive forward the coverage of topics by \Iltis.
    \item Experts in natural language processing for help in building educational tasks for bridging the gap between natural languages and formal modeling.
\end{itemize}

\section*{Acknowledgments}

Many people are contributing to the \Iltis project in various ways. Without them, the system would not exist in its current form. The implementation of the system has been supported by Artur Ljulin, Johannes May, Sebastian Peter, Patrick Roy, and Daniel Sonnabend. The creation of the course material has been supported by Jill Berg, Alicia Gayda, Melanie Schwartz, and Cara Volbracht. Thomas Schwentick and Christopher Spinrath are providing advice for many aspects of the project.

This work was supported by the Deutsche Forschungsgemeinschaft (DFG, German Research Foundation), grant 448468041.

\bibliographystyle{splncs04}
\bibliography{sources}

\end{document}